\documentclass[runningheads]{llncs}
\usepackage{graphicx}
\usepackage{cite}
\usepackage{color}
\usepackage{booktabs}
\usepackage{multicol,listings,multirow}
\usepackage{lipsum}
\usepackage{tikz}
\usepackage{float}
\usepackage{amsmath}
\usepackage{amssymb}
\usepackage{wrapfig}
\usepackage{bbm}
\usepackage{array}
\usepackage{booktabs}
\usepackage{array}
\usepackage{color}
\usepackage{makecell}
\usepackage[colorlinks=true,
            linkcolor=blue,
            filecolor=blue,      
            urlcolor=blue,       
            citecolor=blue
            ]{hyperref}
\usepackage[marginal]{footmisc}
\usepackage[misc]{ifsym}
\usepackage{soul}
\usepackage{url}
\begin{document}
\title{Student Becomes Decathlon Master in Retinal Vessel Segmentation via Dual-teacher Multi-target Domain Adaptation}
\titlerunning{Student Becomes Decathlon Master in Retinal Vessel Segmentation}
%
\author{Linkai Peng\inst{1} \and 
Li Lin\inst{1,2}\and
Pujin Cheng\inst{1} \and
Huaqing He\inst{1} \and
Xiaoying Tang$^{1,3(\textrm{\Letter)}}$}
\authorrunning{L. Peng et al.}
\institute{Department of Electrical and Electronic Engineering, Southern University of Science and Technology, Shenzhen, China\\ 
\email{tangxy@sustech.edu.cn} \and
Department of Electrical and Electronic Engineering, The University of Hong Kong, Hong Kong SAR, China \and
Jiaxing Research Institute, Southern University of Science and Technology, Jiaxing, China}
\maketitle              

\footnote{L. Peng and L. Lin contributed equally to this work.}

\begin{abstract}
\vspace{-0.6cm}
Unsupervised domain adaptation has been proposed recently to tackle the so-called domain shift between training data and test data with different distributions. However, most of them only focus on single-target domain adaptation and cannot be applied to the scenario with multiple target domains. In this paper, we propose RVms, a novel unsupervised multi-target domain adaptation approach to segment retinal vessels (RVs) from multimodal and multicenter retinal images. RVms mainly consists of a style augmentation and transfer (SAT) module and a dual-teacher knowledge distillation (DTKD) module. SAT augments and clusters images into source-similar domains and source-dissimilar domains via B\'ezier and Fourier transformations. DTKD utilizes the augmented and transformed data to train two teachers, one for source-similar domains and the other for source-dissimilar domains. Afterwards, knowledge distillation is performed to iteratively distill different domain knowledge from teachers to a generic student. The local relative intensity transformation is employed to characterize RVs in a domain invariant manner and promote the generalizability of teachers and student models. Moreover, we construct a new multimodal and multicenter vascular segmentation dataset from existing publicly-available datasets, which can be used to benchmark various domain adaptation and domain generalization methods. Through extensive experiments, RVms is found to be very close to the target-trained Oracle in terms of segmenting the RVs, largely outperforming other state-of-the-art methods.

\end{abstract}
\vspace{-0.6cm}
\section{Introduction}

Multimodal ophthalmic images can be effectively employed to extract retinal structures, identify biomarkers, and diagnose diseases. For example, fundus images can depict critical anatomical structures such as the macula, optic disc, and retinal vessels (RVs) \cite{huang2020automated, cheng2021secret}. Optical coherence tomography angiography (OCTA) can efficiently and accurately generate volumetric angiography images \cite{peng2021fargo, lin2021bsda}. These two modalities can both deliver precise representations of vascular structures within the retina, making them popular for the diagnoses of eye-related diseases. These years, advance in the imaging technologies brings new ophthalmic modalities including optical coherence tomography (OCT) \cite{huang1991optical}, widefield fundus \cite{ding2020weakly}, and photoacoustic images \cite{hu2016new}. These new modalities also provide important information of RVs and are likely to enhance the disease diagnosis accuracy. In retinal disease analysis, RV segmentation is a very important pre-requisite. Manual delineation is the most accurate yet highly labor-intensive and time-consuming manner, especially for RV segmentation across multiple modalities. One plausible solution is to manually label RVs on images of a single modality and then transfer the labels to other modalities of interest via advanced image processing and deep learning techniques. However, researches have shown that deep learning models trained on one domain generally perform poorly when tested on another domain with different data distribution, because of domain shift \cite{saenko2010adapting}. This indicates that segmentation models trained on existing modalities will have low generalizability on new modalities. Furthermore, there also exist variations within images of the same modality because of different equipments and imaging settings, posing more challenges to a generalized RV segmentation model for images with cross-center or cross-modality domain shift.

To address this issue, domain adaptation has been explored, among which unsupervised domain adaptation (UDA) has gained the most favor in recent years. For instance, Javanmardi et al. \cite{javanmardi2018domain} combined U-net \cite{ronneberger2015u} with a domain discriminator and introduced adversarial training based on DAAN \cite{ganin2016domain}. Wang et al. \cite{wang2019patch} adopted this design to joint optic disc and cup segmentation from fundus images. These two methods focus on domain shift with small variations (i.e., cross-center domain shift). To tackle domain shift with large variations (i.e., cross-modality domain shift), Cai et al. \cite{cai2019towards} incorporated CycleGAN \cite{zhu2017unpaired} into its own network and employed a shape consistency loss to ensure a correct translation of semantics in medical images. Dou et al. \cite{dou2019pnp} designed a cross-modality UDA framework for cardiac MR and CT image segmentation by performing DA only at low-level layers, under the assumption that domain shift mainly lies in low-level characteristics. Zhang et al. \cite{zhang2019noise} presented a Noise Adaptation Generative Adversarial Network for RV segmentation and utilized a style discriminator enforcing the translated images to have the same noise patterns as those in the target domain. Peng et al. \cite{peng2022unsupervised} employed disentangled representation learning to disentangle images into content space and style space to extract domain-invariant features. These approaches are nevertheless designed to handle only a specific type of domain shift and may have limitations when applied to multimodal and multicenter (\emph{m}\&\emph{m}) ophthalmic images.

We here propose a novel unsupervised multi-target domain adaptation (MTD\\A) approach, RVms, to segment RVs from \emph{m}\&\emph{m} retinal images. RVms mainly consists of two components: a style augmentation and transfer (SAT) module and a dual-teacher knowledge distillation (DTKD) module. SAT uses B\'ezier transformation to augment source images into diverse styles. By adjusting the interpolation rate, source-style images are gradually transferred to target styles via Fourier Transform. We then cluster images into source-similar domains $D_{sim}$ and source-dissimilar domains $D_{dis}$ according to the vessel-background relative intensity information. DTKD utilizes the augmented and transformed data to train two teachers, one for $D_{sim}$ and the other for $D_{dis}$. Then knowledge distillation \cite{hinton2015distilling} is performed to iteratively distill different domain knowledge from teachers to a generic student. In addition, the local relative intensity transformation (LRIT) is employed to characterize RVs in a domain invariant manner and to improve the generalizability of teachers and student models. 

The main contributions of this work are four-fold: (1) We propose a novel MTDA framework for RV segmentation from \emph{m}\&\emph{m} retinal images. To the best of our knowledge, this is the first work that explores unsupervised MTDA for medical image segmentation across both modalities and centers. (2) MTDA features itself with style augmentation, style transfer, and dual-teacher knowledge distillation, which effectively tackles both cross-center and cross-modality domain shift. (3) Extensive comparison experiments are conducted, successfully identifying the proposed pipeline's superiority over representative state-of-the-art (SOTA) methods. (4) We construct a new \emph{m}\&\emph{m} vascular segmentation dataset from existing publicly-available datasets, which can be used to benchmark various domain adaptation and domain generalization methods. Our code and dataset are publicly available at \url{https://github.com/lkpengcs/RVms}.

\vspace{-0.2cm}
\section{Methodology}

\begin{figure}[!t]
	\centering
	\vspace{-0.2cm}
	\setlength{\belowcaptionskip}{-0.7cm}   
	\centerline{\includegraphics[width=11cm]{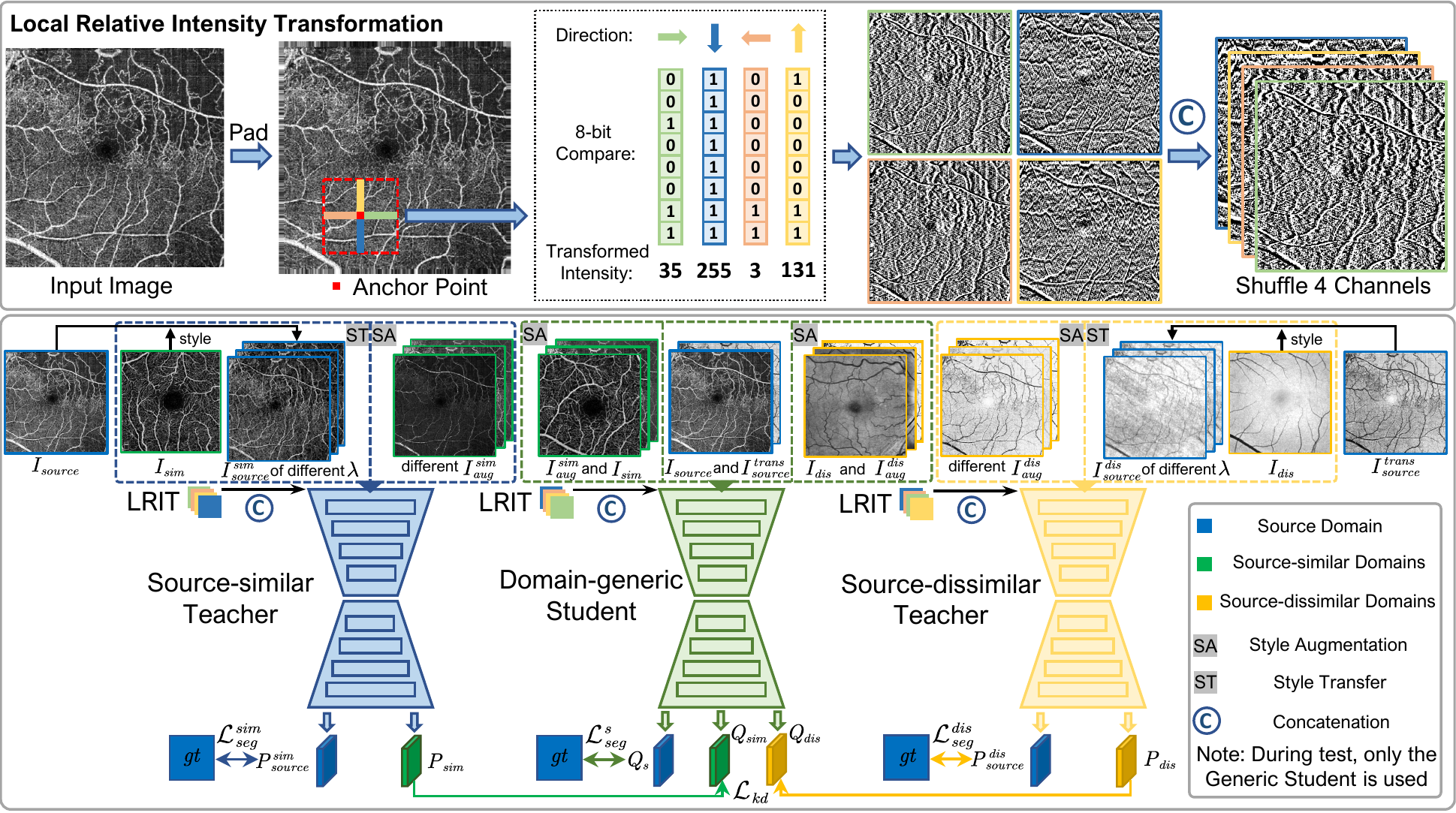}}
	\caption{Schematic demonstration of the architecture of our RVms framework. The upper part represents the LRIT module and the lower part represents the DTKD module. $P(\cdot)$ is the prediction of teachers and $Q(\cdot)$ is the prediction of the student.}\medskip
	\vspace{-0.15cm}
	\label{model}
\end{figure}

The proposed RVms framework is shown in Fig. \ref{model}, which consists of a SAT module and a DTKD module.

\vspace{-0.2cm}
\subsection{Definition}

We define the source domain as $S=\left\{x_{i}^{s}, y_{i}^{s}\right\}_{i=1}^{N^{s}}$, where $x_i^s$ is the \emph{i}-th input image, $y_i^s$ is the corresponding RV segmentation label, and $N^s$ is the total number of source domain images. The target domains are denoted as $T=\left\{T_{1}, T_{2}, \ldots T_{n}\right\}$, where $T_{t}=\left\{x_{j}^{t}, y_{j}^{t}\right\}_{j=1}^{N^{t}}$, $x_j^t$ is the \emph{j}-th image of the \emph{t}-th target domain, $y_j^t$ is the corresponding RV segmentation label, and $N^t$ is the total number of images from the \emph{t}-th target domain. From SAT, we obtain images in source-similar domains $D_{sim}$ and source-dissimilar domains $D_{dis}$. Then in DTKD, a source-similar teacher $T_{sim}$ and a source-dissimilar teacher $T_{dis}$ are separately trained. Finally, we employ $T_{sim}$ and $T_{dis}$ to distill knowledge to a generic student $S_g$.

\vspace{-0.2cm}
\subsection{Style Augmentation and Transfer}

\subsubsection{Style Augmentation.} Retinal images are either grayscale or can be converted to grayscale and the content across different images is generally similar; the difference mainly lies in the intensity and the clarity of tiny vessels. Besides, the style in different modalities is also a key factor that leads to domain shift, especially for small-variation domain shift. Thus, we follow the work of \cite{zhou2019models} and utilize a non-linear transformation via the monotonic and smooth B\'ezier Curve function. It is a one-to-one mapping that assigns each pixel a new and unique value. We generate the B\'ezier Curve from two end points ($P_0$ and $P_3$) and two control points ($P_1$ and $P_2$), which is defined as
\begin{equation}
\vspace{-0.05cm}
B(t)=(1-t)^{3} P_{0}+3(1-t)^{2} t P_{1}+3(1-t) t^{2} P_{2}+t^{3} P_{3},\ t \in[0,1],
\vspace{-0.05cm}
\end{equation}
where $t$ is a fractional value along the length of the line. We set $P_0 = (0,0)$ and $P_3 = (1,1)$ to get source-similar augmentations and the opposite to get source-dissimilar augmentations. The $x$-axis coordinates and $y$-axis coordinates of $P_1$ and $P_2$ are randomly selected from the interval $(0,1)$.

\vspace{-0.3cm}
\subsubsection{Style Transfer.} 
\begin{figure}[!t]
	\centering
	\vspace{-0.1cm}
	\setlength{\belowcaptionskip}{-0.7cm}   
	\centerline{\includegraphics[width=10.8cm]{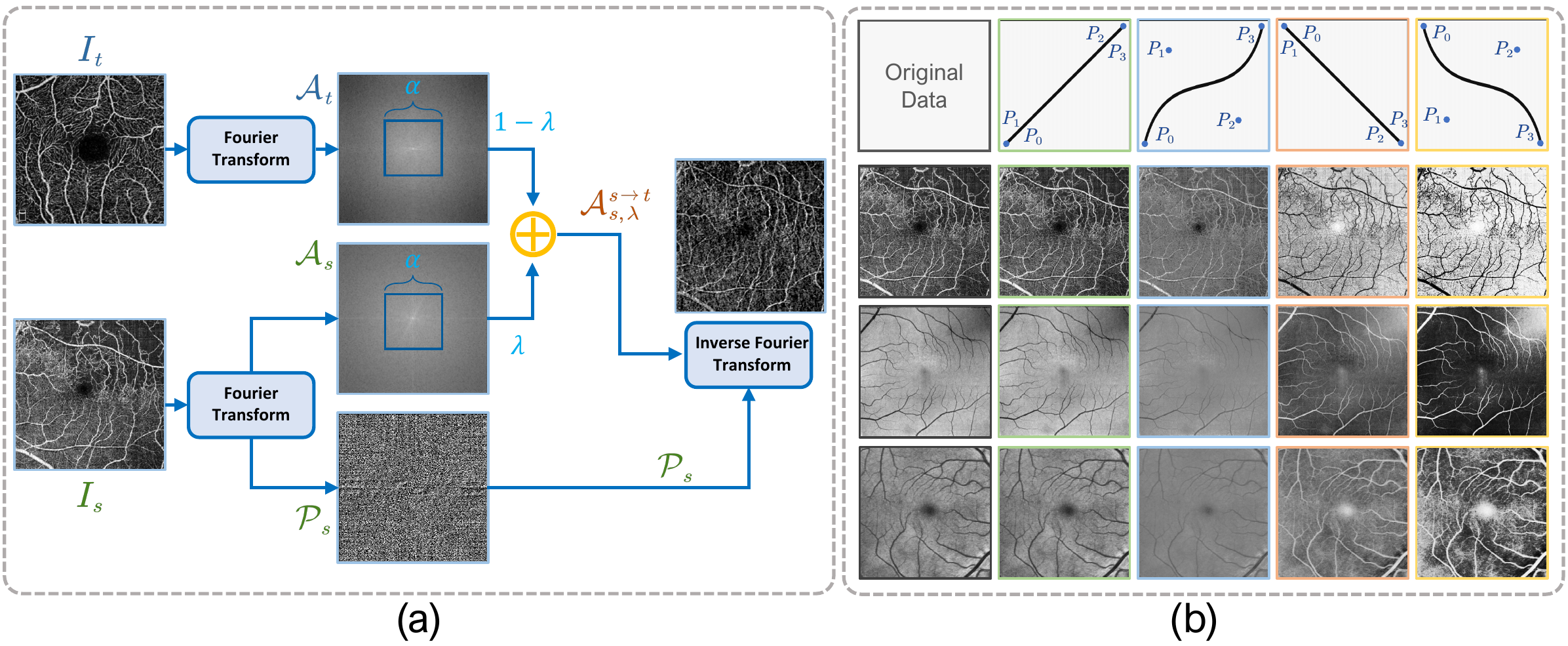}}
	\vspace{-0.3cm}
	\caption{Schematic demonstration of our proposed SAT module. Panel (a) is the style transfer module and Panel (b) is the style augmentation module.}\medskip
	\vspace{-0.3cm}
	\label{submodel}
\end{figure}

In addition to grayscale differences, different modalities also differ in the image style, such as vessel shape and existence of other anatomical landmarks. To imitate the diverse appearance of different target domains, we adopt Fourier Transform to progressively translate the style of source domain to those of target domains. Given a source image $x_i^s$ and a randomly selected target image $x_j^t$ from target domain $T_t$, we first perform Fourier Transform on both images to get amplitude spectrums $\mathcal{A}_s, \mathcal{A}_t$ and phase spectrums $\mathcal{P}_s, \mathcal{P}_t$ \cite{liu2021feddg, yang2020fda}. Then we use a binary mask $\mathcal{M}=\mathbbm{1}_{(h, w) \in[-\alpha H: \alpha H,-\alpha W: \alpha W]}$ to extract the central regions of $\mathcal{A}_s$ and $\mathcal{A}_t$ and combine them with an interpolation rate $\lambda$. In this way the contributions of $\mathcal{A}_s$ and $\mathcal{A}_t$ to the synthesized image can be adjusted.
\begin{equation}
\vspace{-0.4cm}
\label{eq2}
\mathcal{A}_{s, \lambda}^{s \rightarrow t}=\left((1-\lambda) \mathcal{A}_s+\lambda \mathcal{A}_t\right) * \mathcal{M}+\mathcal{A}_s *(1-\mathcal{M}).   
\vspace{0.3cm}
\end{equation}
Finally, we perform inverse Fourier Transform to obtain an image with content from source domain and style from both source and target domains (Eq. \ref{eq3}). Detailed illustration of SAT is shown in Fig. \ref{submodel}.

\begin{equation}
\vspace{-0.3cm}
\label{eq3}
x_{i, \lambda}^{s \rightarrow t}=\mathcal{F}^{-1}\left(\mathcal{A}_{s, \lambda}^{s \rightarrow t}, \mathcal{P}_s\right).
\end{equation}

\vspace{-0.3cm}
\subsection{Dual-Teacher Knowledge Distillation}

\subsubsection{Local Relative Intensity Transformation.} In our situation, despite that retinal vessels are in different modalities and multi-scaled, they always have a consistent relationship with the background in grayscale intensity. This relationship can be used to depict vessels in a domain-invariant manner. We employ LRIT to extract RV features by taking advantage of this vessel-background relationship. Inspired by \cite{ojala2002multiresolution, shi2022local}, each pixel serves as an anchor point and the intensity values of the adjacent eight pixels are compared to generate a new value for the anchor point through the formula in Eq. \ref{eq4}. Adjacent pixels in all four directions (i.e., up, down, left, and right) are compared separately, resulting in four transformed images.
\begin{equation}
\label{eq4}
\vspace{-0.3cm}
\begin{split}
V_{new}(a)&=\sum_{i=1}^{8}c\left(V(a)-V\left(n_{i}\right)\right) \times 2^{i},\\
c(x)&=\left\{\begin{array}{l}
1, \quad \text { if } x>0 \\
0, \text { otherwise },
\end{array}\right.
\end{split}
\end{equation}
where $V(\cdot)$ is the intensity value and $n_i$ is the \emph{i}-th adjacent pixel of the anchor point. For pixels near edges, we use edge padding to ensure there are eight neighboring pixels in each direction. The four transformed images are shuffled and concatenated as external channels for both teachers and student models.

\vspace{-0.3cm}
\subsubsection{Knowledge Distillation.} The dual teachers $T_{sim}$ and $T_{dis}$ are respectively trained with augmented and style-transferred images in $D_{sim}$ and $D_{dis}$. The source images, style augmented images, and style transferred images are used as inputs and the corresponding labels are used for supervision via Dice losses $\mathcal{L}_{seg}^{sim}$ and $\mathcal{L}_{seg}^{dis}$. When the two teachers are trained to converge (after $\tau$ epochs), we start the knowledge distillation process. The domain-generic student model $S_g$ is supervised with the outputs of $T_{sim}$ and $T_{dis}$, each being responsible for one type of the target domains (source-similar or source-dissimilar). Specifically, we adopt a cross-entropy loss $\mathcal{L}_{kd}$ to minimize the distribution differences between the corresponding outputs of the teachers and student models (i.e., $P_{sim}$ and $Q_{sim}$, $P_{dis}$ and $Q_{dis}$). The groundtruth is also used to supervise the training on the source domain through a Dice loss $\mathcal{L}_{seg}^{s}$. The total loss function is
\begin{equation}
\begin{aligned}
\mathcal{L}_{DTKD }=\left\{\begin{array}{cc}
\mathcal{L}_{seg}^{sim} + \mathcal{L}_{seg}^{dis}, & \text { epoch } \leqslant \tau \\
\mathcal{L}_{kd} + \mathcal{L}_{seg}^{s} + \mathcal{L}_{seg}^{sim} + \mathcal{L}_{seg}^{dis}, & \text { epoch }>\tau.
\end{array}\right.
\end{aligned}
\end{equation}

\section{Experiments and Results}

\vspace{-0.1cm}
\subsection{Dataset}

We construct a new dataset named mmRV, consisting of five domains from five publicly available datasets \cite{li2020ipn, ma2020rose, staal2004ridge, budai2013robust, ding2020weakly}. The details are shown in Table \textcolor{blue}{A1} of the appendix. For OCTA-500, we resize each image to 384 $\times$ 384 and discard samples with severe quality issues. For fundus images, we crop out the micro-vascular region surrounding the macula to avoid interference from other unrelated structures such as the optic disc and only focus on vessels; the capillaries near the macula are mostly multi-scaled and difficult to segment \cite{lin2020blu}. We also apply Contrast Limited Adaptive Histogram Equalization to fundus images as preprocessing. For PRIME-FP20, we augment each image for four times.

\vspace{-0.2cm}
\subsection{Experimental Setting}

We train our RVms framework on the newly-constructed mmRV dataset. We conduct two sets of MTDA experiments respectively using DRIVE and OCTA-500 as the source domain because the image modalities in those two datasets are the most commonly used in clinical practice and those two datasets are most well-annotated. All compared methods and RVms are implemented with Pytorch using NVIDIA RTX 3090 GPUs. For both teachers and student models in RVms, we use the Adam optimizer with a learning rate of $1\times10^{-3}$. The number of epochs $\tau$ for training the dual teachers is set to be 200 and then we co-train the teachers with the student for another 400 epochs. $\alpha$ is set to 0.2 and $\lambda$ is randomly selected in $(0,1)$. During testing, the target domain images are directly inputted to $S_g$ to get the corresponding predictions. We use ResNet34 \cite{he2016deep} with ImageNet pretrained initialization as the encoder in a modified U-net \cite{ronneberger2015u} architecture.

\vspace{-0.2cm}
\subsection{Results}

All methods are evaluated using two metrics, i.e., Dice[\%] and 95\% Hausdorff Distance (HD[px]), the results of which are tabulated in Table \ref{result_drive}. We compare RVms with two recently-developed SOTA DA/MTDA models, namely ADVENT \cite{vu2019advent} and Multi-Dis \cite{saporta2021multi}. Note that \cite{vu2019advent} is trained with mixed target domains. We also compare with a domain generalization method Dofe \cite{wang2020dofe}, which leaves one out as the target domain and uses multiple source-similar domains as the source domains. The results from the dual teachers $T_{sim}$ and $T_{dis}$ are also reported. \textbf{Source Only} means the model is trained with source domain data only. \textbf{Oracle} means the model is trained and tested on the specific target domain. Our method achieves Dice scores that are about 38\% and 13\% higher than the second best method when using OCTA (OCTA-500) and fundus image (DRIVE) as the source domain. It is evident that our proposed RVms delivers superior RV segmentation performance when encountering both cross-modality and cross-center domain shift. Besides, it achieves the highest average Dice score.

\begin{table}[!h]
\vspace{-0.5cm}
\caption{Quantitative evaluations of different methods. \textbf{Bold} and \underline{underlined} numbers respectively denote the best and second-best results.}
\vspace{-0.2cm}
\label{result_drive}
\centering
\resizebox{0.95\textwidth}{!}{
\begin{tabular}{c|cccccccccccc}
\Xhline{1.2pt}
 & \multicolumn{12}{c}{Source Domain: OCTA (OCTA-500)} \\ \Xhline{1.2pt}
Modalities & \multicolumn{2}{c}{\begin{tabular}[c]{@{}c@{}}Fundus Image\\ (DRIVE)\end{tabular}} & \multicolumn{2}{c}{\begin{tabular}[c]{@{}c@{}}OCTA\\ (ROSE)\end{tabular}} & \multicolumn{2}{c}{\begin{tabular}[c]{@{}c@{}}OCT\\ (OCTA-500)\end{tabular}} & \multicolumn{2}{c}{\begin{tabular}[c]{@{}c@{}}Fundus Image\\ (HRF)\end{tabular}} & \multicolumn{2}{c}{\begin{tabular}[c]{@{}c@{}}UWF Fundus\\ (PRIME-FP20)\end{tabular}} & \multicolumn{2}{c}{Average} \\ \Xhline{1.2pt}
Metrics & Dice $\uparrow$ & HD $\downarrow$ & Dice $\uparrow$ & HD $\downarrow$  & Dice $\uparrow$ & HD $\downarrow$  & Dice $\uparrow$ & HD $\downarrow$  & Dice $\uparrow$ & HD $\downarrow$  & Dice $\uparrow$ & HD $\downarrow$ \\ \Xhline{1.2pt}
Source Only & 2.18 & 178.06 & 52.35 & 25.52 & 5.70 & 62.89 & 0.43 & 189.60 & 1.33 & 180.42 & 12.40 & 127.30 \\
Dofe & 24.03 & 27.91 & 48.52 & 21.21 & 34.41 & 21.76 & 24.67 & 25.63 & 30.37 & 24.75 & 32.40 & 24.25 \\
ADVENT & 1.46 & 52.53 & 52.86 & \underline{12.80} & 4.91 & 44.48 & 0.37 & 44.16 & 1.01 & \underline{20.11} & 12.12 & 34.81 \\
Multi-Dis & 1.45 & 39.65 & \underline{58.98} & \textbf{10.29} & 5.07 & 30.77 & 0.31 & 46.91 & 1.03 & 52.09 & 13.37 & 35.94 \\
$T_{sim}$ & 5.35 & 36.18 & 53.04 & 23.27 & 2.92 & 32.89 & 2.90 & 30.75 & 5.07 & 37.08 & 13.85 & 39.27 \\
$T_{dis}$ & \textbf{72.77} & \textbf{10.77} & 5.68 & 19.52 & \textbf{77.76} & \textbf{15.60} & \underline{72.90} & \textbf{12.12} & \textbf{76.49} & \textbf{12.57} & \underline{49.73} & \textbf{14.16} \\
Ours & \underline{72.64} & \underline{17.72} & \textbf{60.80} & 13.49 & \underline{72.81} & \underline{24.70} & \textbf{74.03} & \underline{12.97} & \underline{74.54} & 21.09 & \textbf{70.96} & \underline{17.99} \\
Oracle & \textcolor{gray}{67.72} & \textcolor{gray}{9.05} & \textcolor{gray}{71.43} & \textcolor{gray}{12.14} & \textcolor{gray}{82.68} & \textcolor{gray}{11.13} & \textcolor{gray}{72.09} & \textcolor{gray}{12.67} & \textcolor{gray}{77.80} & \textcolor{gray}{9.33} & \textcolor{gray}{74.34} & \textcolor{gray}{10.86}  \\ \Xhline{1.2pt}
 & \multicolumn{12}{c}{Source Domain: Fundus Image (DRIVE)} \\ \Xhline{1.2pt}
Modalities & \multicolumn{2}{c}{\begin{tabular}[c]{@{}c@{}}OCTA\\ (OCTA-500)\end{tabular}} & \multicolumn{2}{c}{\begin{tabular}[c]{@{}c@{}}OCTA\\ (ROSE)\end{tabular}} & \multicolumn{2}{c}{\begin{tabular}[c]{@{}c@{}}OCT\\ (OCTA-500)\end{tabular}} & \multicolumn{2}{c}{\begin{tabular}[c]{@{}c@{}}Fundus Image\\ (HRF)\end{tabular}} & \multicolumn{2}{c}{\begin{tabular}[c]{@{}c@{}}UWF Fundus\\ (PRIME-FP20)\end{tabular}} & \multicolumn{2}{c}{Average} \\ \Xhline{1.2pt}
Metrics & Dice $\uparrow$ & HD $\downarrow$ & Dice $\uparrow$ & HD $\downarrow$  & Dice $\uparrow$ & HD $\downarrow$  & Dice $\uparrow$ & HD $\downarrow$  & Dice $\uparrow$ & HD $\downarrow$  & Dice $\uparrow$ & HD $\downarrow$ \\ \Xhline{1.2pt}
Source Only & 11.39 & 21.81 & 17.69 & \textbf{10.24} & 59.24 & 19.77 & 58.77 & \textbf{10.38} & 64.87 & 14.92 & 42.39 & \textbf{15.42} \\
Dofe & 30.94 & 22.72 & 38.39 & 11.48 & \underline{62.69} & 19.13 & \underline{60.28} & \underline{12.86} & \underline{65.46} & \textbf{13.88} & \underline{51.55} & 16.01 \\
ADVENT & 22.38 & 47.40 & 20.36 & 33.88 & 47.32 & 42.66 & 53.15 & 34.59 & 52.86 & 42.68 & 39.21 & 40.24 \\
Multi-Dis & 32.75 & 21.30 & 33.73 & \underline{11.16} & 52.42 & 35.79 & 59.46 & 28.22 & 60.62 & 38.54 & 47.80 & 27.00 \\
$T_{sim}$ & 12.05 & 19.58 & 14.32 & 23.08 & 59.11 & 21.19 & 59.15 & 15.37 & 64.96 & 17.69 & 41.91 & 19.38 \\
$T_{dis}$ & \underline{64.44} & \underline{18.51} & \textbf{63.81} & 16.13 & 11.01 & \textbf{14.63} & 6.87 & 25.28 & 10.08 & 24.69 & 31.24 & 19.84 \\
Ours & \textbf{68.93} & \textbf{15.67} & \underline{61.31} & 16.05 & \textbf{64.35} & \underline{18.93} & \textbf{63.18} & 13.50 & \textbf{69.44} & \underline{14.04} & \textbf{65.44} & \underline{15.63}\\
Oracle & \textcolor{gray}{88.35} & \textcolor{gray}{4.21} & \textcolor{gray}{71.43} & \textcolor{gray}{12.14} & \textcolor{gray}{82.68} & \textcolor{gray}{11.13} & \textcolor{gray}{72.09} & \textcolor{gray}{12.67} & \textcolor{gray}{77.80} & \textcolor{gray}{9.33} & \textcolor{gray}{78.47} & \textcolor{gray}{9.89} \\ \Xhline{1.2pt}
\end{tabular}}
\vspace{-0.6cm}
\end{table}
 
Representative visualization results are illustrated in Fig. \ref{result} and Fig. \textcolor{blue}{A1} of the appendix. We observe that all compared methods fail in many cases while our framework is very close to the Oracle. Apparently, our framework achieves superior performance when tested on all target domains.

\begin{figure}[!t]
	\centering
	\vspace{-0.15cm}
	\setlength{\belowcaptionskip}{-0.7cm}   
	\centerline{\includegraphics[width=10.6cm
	]{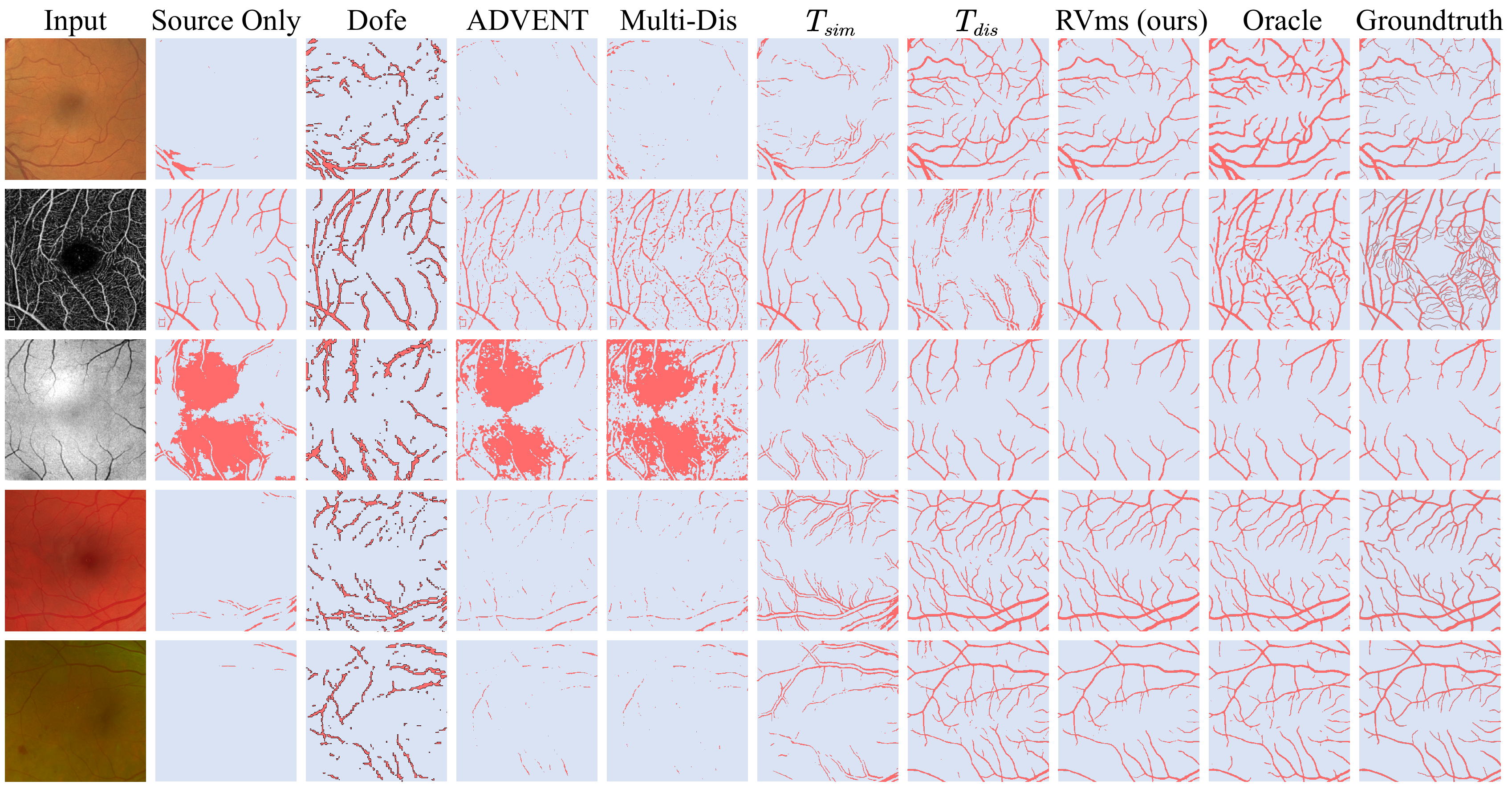}}
	\vspace{-0.3cm}
	\caption{Representative RV segmentation results using the OCTA images from OCTA-500 as the source domain. The target domains are Fundus Image (DRIVE), OCTA (ROSE), OCT (OCTA-500), Fundus Image (HRF), and UWF Fundus (PRIME-FP20) from top to bottom.}\medskip
	\vspace{0.6cm}
	\label{result}
\end{figure}

To evaluate the effectiveness of several key components in RVms, we conduct ablation studies. We compare with the proposed RVms without style augmentation (w/o sa), without style transfer (w/o st), without LRIT (w/o LRIT). We also compare with training a single model with images in $D_{sim}$ and $D_{dis}$ without knowledge distillation (w/o KD) and employing only one teacher model (w/o DT). The results are shown in Table \ref{ablation_result_octa} and Table \textcolor{blue}{A2} of the appendix. The performance degrades when removing any component in the framework, in terms of the average Dice score.

\begin{table}[!h]
\vspace{-0.2cm}
\caption{Ablation analysis results for several key components in our proposed framework using OCTA (OCTA-500) as the source domain. \textbf{Bold} and \underline{underlined} numbers respectively denote the best and second-best results.}
\vspace{-0.2cm}
\label{ablation_result_octa}
\centering
\resizebox{0.96\textwidth}{!}{
\begin{tabular}{c|cccccccccccc}
\Xhline{1.2pt}
 & \multicolumn{12}{c}{Source Domain: OCTA (OCTA-500)} \\ \Xhline{1.2pt}
Modalities & \multicolumn{2}{c}{\begin{tabular}[c]{@{}c@{}}Fundus Image\\ (DRIVE)\end{tabular}} & \multicolumn{2}{c}{\begin{tabular}[c]{@{}c@{}}OCTA\\ (ROSE)\end{tabular}} & \multicolumn{2}{c}{\begin{tabular}[c]{@{}c@{}}OCT\\ (OCTA-500)\end{tabular}} & \multicolumn{2}{c}{\begin{tabular}[c]{@{}c@{}}Fundus Image\\ (HRF)\end{tabular}} & \multicolumn{2}{c}{\begin{tabular}[c]{@{}c@{}}UWF Fundus\\ (PRIME-FP20)\end{tabular}} & \multicolumn{2}{c}{Average} \\ \Xhline{1.2pt}
Metrics & Dice $\uparrow$ & HD $\downarrow$ & Dice $\uparrow$ & HD $\downarrow$  & Dice $\uparrow$ & HD $\downarrow$  & Dice $\uparrow$ & HD $\downarrow$  & Dice $\uparrow$ & HD $\downarrow$  & Dice $\uparrow$ & HD $\downarrow$ \\ \Xhline{1.2pt}
Source Only & 2.18 & 178.06 & 52.35 & 25.52 & 5.70 & 62.89 & 0.43 & 189.60 & 1.33 & 180.42 & 12.40 & 127.30 \\
w/o sa & 62.74 & 24.82 & 48.25 & 26.51 & 60.47 & 30.50 & 60.48 & 30.89 & 60.17 & 29.91 & 58.42 & 26.73 \\
w/o st & 70.19 & 21.58 & 41.95 & 27.29 & 71.59 & 24.81 & 70.86 & 18.19 & 69.57 & 23.33 & 55.94 & 23.04 \\
w/o LRIT & \underline{71.86} & \textbf{13.35} & \underline{56.65} & \underline{19.93} & 71.60 & \underline{18.84} & 71.54 & \textbf{10.28} & \underline{73.83} & \textbf{15.17} & \underline{69.09} & \textbf{15.31} \\
w/o KD & 69.73 & \underline{14.78} & 13.08 & 20.57 & 71.07 & \textbf{18.83} & 70.53 & 16.75 & 71.96 & \underline{18.54} & 59.27 & \underline{17.89} \\
w/o DT & 70.08 & 16.06 & 43.20 & 26.12 & \underline{72.11} & 20.91 & \underline{72.21} & 13.85 & 72.25 & 19.64 & 65.97 & 19.31 \\
Ours & \textbf{72.64} & 17.72 & \textbf{60.80} & \textbf{13.49} & \textbf{72.81} & 24.70 & \textbf{74.03} & \underline{12.97} & \textbf{74.54} & 21.09 & \textbf{70.96} & 17.99 \\
Oracle & \textcolor{gray}{67.72} & \textcolor{gray}{9.05} & \textcolor{gray}{71.43} & \textcolor{gray}{12.14} & \textcolor{gray}{82.68} & \textcolor{gray}{11.13} & \textcolor{gray}{72.09} & \textcolor{gray}{12.67} & \textcolor{gray}{77.80} & \textcolor{gray}{9.33} & \textcolor{gray}{74.34} & \textcolor{gray}{10.86}  \\
\Xhline{1.2pt}
\end{tabular}}
\vspace{-0.5cm}
\end{table}

\vspace{-0.3cm}
\section{Conclusion}

\vspace{-0.3cm}
In this paper, we proposed and validated a novel framework for unsupervised multi-target domain adaptation in retinal vessel segmentation. We used style augmentation and style transfer to generate source-similar images and source-dissimilar images to improve the robustness of both teachers and student models. We also conducted knowledge distillation from dual teachers to a generic student, wherein a domain invariant method named LRIT was utilized to facilitate the training process. Another contribution of this work is that we constructed a new dataset called mmRV from several public datasets, which can be used as a new benchmark for DA and domain generalization. Through extensive experiments, our proposed RVms was found to largely outperform representative SOTA MTDA methods, in terms of RV segmentation from different modalities.

\vspace{-0.3cm}
\section{Acknowledgements}

\vspace{-0.3cm}
This study was supported by the Shenzhen Basic Research Program (JCYJ20190\\809120205578); the National Natural Science Foundation of China (62071210); the Shenzhen Science and Technology Program (RCYX20210609103056042); the Shenzhen Basic Research Program (JCYJ20200925153847004); the Shenzhen Science and Technology Innovation Committee (KCXFZ2020122117340001).

%
%
%
%
\newpage

\end{document}